\documentclass[letter,times,twocolumn,tighten]{aastex62}
\hypersetup{linkcolor=red,citecolor=blue,filecolor=cyan,urlcolor=blue}
\usepackage{amsmath,amstext}
\usepackage[figure,figure*]{hypcap}
\usepackage{commath}

\definecolor{oxfordblue}{rgb}{0.0, 0.13, 0.28}
\definecolor{firebrick1}{rgb}{0.698, 0.133, 0.133}

\shorttitle{Migrating mini-Neptune in HD~169142}
\shortauthors{P\'erez et al.}
\begin{document}

\title{DUST UNVEILS THE FORMATION OF A MINI-NEPTUNE PLANET IN A PROTOPLANETARY RING}

\correspondingauthor{Sebasti\'an P\'erez}
\email{sebastian.astrophysics@gmail.com}
\author[0000-0003-2953-755X]{Sebasti\'an P\'erez}

\affil{Universidad de Santiago de Chile, Av. Libertador Bernardo O'Higgins 3363, Estaci\'on Central, Santiago}
\affil{Departamento de F\'isica, Universidad de Santiago de Chile, Av. Ecuador 3493, Estaci\'on Central, Santiago}
\affil{Departamento de Astronom\'ia, Universidad de Chile, Casilla 36-D, Santiago}

\author[0000-0002-0433-9840]{Simon Casassus}
\affiliation{Departamento de Astronom\'ia, Universidad de Chile, Casilla 36-D, Santiago}

\author[0000-0002-2672-3456]{Cl\'ement Baruteau}
\affiliation{IRAP, Universit\'e de Toulouse, CNRS, UPS, Toulouse, France}

\author[0000-0001-9290-7846]{Ruobing Dong}
\affiliation{Department of Physics \& Astronomy, University of Victoria, Victoria, BC, V8P 1A1}

\author[0000-0001-5073-2849]{Antonio Hales}
\affiliation{National Radio Astronomy Observatory, 520 Edgemont Road, Charlottesville, VA 22903-2475.}
\affiliation{Atacama Large Millimeter/Submillimeter Array, Joint ALMA Observatory, Alonso de C\'ordova 3107, Vitacura 763-0355, Santiago}

\author[0000-0002-2828-1153]{Lucas Cieza}
\affiliation{N\'ucleo de Astronom\'ia, Facultad de Ingenier\'a y Ciencias, Universidad Diego Portales, Av. Ejercito 441, Santiago, Chile}

\begin{abstract}
  Rings and radial gaps are ubiquitous in protoplanetary disks, yet
  their possible connection to planet formation is currently subject
  to intense debates. In principle, giant planet formation leads to
  wide gaps which separate the gas and dust mass reservoir in the
  outer disk, while lower mass planets lead to shallow gaps which are
  manifested mainly on the dust component. We used the Atacama Large
  Millimeter/submillimeter Array (ALMA) to observe the star
  HD\,169142, host to a prominent disk with deep wide gaps that sever
  the disk into inner and outer regions.  The new ALMA high resolution
  images allow for the outer ring to be resolved as three narrow
  rings.  The HD\,169142 disk thus hosts both the wide gaps trait of
  transition disks and a narrow ring system similar to those observed
  in the TW~Hya and HL~Tau systems. The mass reservoir beyond a deep
  gap can thus host ring systems. The observed rings are narrow in
  radial extent (width/radius of 1.5/57.3, 1.8/64.2 and 3.4/76.0,
  in~{\sc au}) and have asymmetric mutual separations: the first and
  middle ring are separated by 7~{\sc au} while the middle and
  outermost ring are distanced by $\sim$12~{\sc au}. Using
  hydrodynamical modeling we found that a simple explanation,
  involving a single migrating low mass planet (10~M$_\oplus$),
  entirely accounts for such an apparently complex phenomenon.  Inward
  migration of the planet naturally explains the ring's asymmetric
  mutual separation. The isolation of HD\,169142's outer rings thus
  allows a proof of concept to interpret the detailed architecture of
  the outer region of protoplanetary disks with low mass planet
  formation of mini-Neptune's size, i.e. as in the protosolar nebula.
\end{abstract}


\section{Introduction} \label{sec:intro}

Planet formation theories describe two main pathways for the formation
of protoplanet embryos: gravitational instability followed by
fragmentation, which forms gas giant planets, or (sub-)stellar mass
companions \citep[e.g.,][]{Boss1997,Kratter2010}; and bottom-up growth
by core accretion, which forms terrestrial and icy planets, or the
cores of giant planets \citep[e.g.,][]{Pollack1996}. However, the
timescale of core accretion increases dramatically with stellocentric
distance due to the longer dynamical timescale and low surface
densities at large radii \citep[e.g.,][]{Goldreich2004}, which poses a
problem for the formation of planets before the protoplanetary disk
dissipates.  Mechanisms such as pebble accretion \citep{Ormel2010} are
thought to alleviate the tension required to form the cores of
Jupiter, Saturn, and the icy giants in the Solar System (all within
30\,{\sc au}) and the lifetime of the protosolar disk
\citep{Lambrechts2012}.

The formation and evolution of pebbles and planetesimals, and hence
planetary cores, relies upon radial pressure bumps that curb the
catastrophic drift of solids towards the star due to aerodynamic drag
\citep[e.g.,][]{Pinilla2012}.  These radial dust traps have been
associated with the ring systems observed in disks in the thermal
emission from cold dust grains at radio wavelengths. Forming planets
are thought to open gaps in the dust distribution, and the masses of
such putative protoplanets are commonly inferred from the gap's depth
and width \citep{Rosotti2016, DongFung2017}. It is frequent to
associate one planet per gap \citep{Dipierro2015, Mentiplay2018,
  Clarke2018}, multiple planets to one (wide) gap \citep{Dong2015},
and one planet to multiple (narrow) gaps \citep[e.g., a 0.1$M_{\rm
    Jup}$ planet can explain the multiple gaps in
  AS~209,][]{Zhang2018}.

In the so-called `gapped disks', sometimes called transitional disks,
dust is evacuated from central cavities or deep radial gaps, resulting
in one or two bright rings. Accreting giant planets of masses
comparable to or larger than that of Jupiter are able to open deep
gaps in the gas and accumulate dust into an outer ring
\citep{Crida2006, Pinilla2012, Dipierro2016}. By contrast, in disks
without a deep gap, dust accumulation in mild pressure maxima produces
sequences of several {\em shallow} rings \citep[e.g., the TW~Hya
  disk,][]{Andrews2016}.  Clear signs of low-mass planet formation, as
hinted by these fine shallow rings \citep{Dong2017, Dong2018}, have
been absent in the outer regions of `gapped disks'.

The 1.7~$M_\odot$ star HD\,169142 \citep[estimated age
  6$^{+6}_{-3}$~Myr,][]{Grady2007} hosts a prominent disk with deep
wide gaps imaged in the near-infrared \citep{Quanz2013, Momose2015,
  Pohl2017, Monnier2017, Ligi2018, Bertrang2018}, mid-infrared
\citep{Honda2012}, 1.3 millimeter \citep{Fedele2017}, and centimetre
\citep{Osorio2014, Macias2018} wavelengths.  Circum-planetary features
related to giant protoplanets have tentatively been reported in
HD\,169142 from radio/IR imaging \citep{Reggiani2014, Biller2014,
  Osorio2014, Gratton2019}, but the nature of these signals remains to
be confirmed \citep{Ligi2018}.

In this paper, we present new 1.3~mm observations of the dust
structures around HD\,169142 at $\sim$2~{\sc au} resolution
(Sec.~\ref{sec:obs}). In these new observations, the outer disk shows
an intricate system of fine rings and narrow gaps (described in
Sec.~\ref{sec:res}). Using hydrodynamical modeling
(Sec.~\ref{sec:model}), a simple connection between the fine ring
structure and a single, {\em migrating}, mini-Neptune protoplanet can
be made, which we present and discuss in Sec.~\ref{sec:model2}.
Implications are discussed in Sec.~\ref{sec:summary}.

\section{Observations and data reduction} \label{sec:obs}
 
We obtained 1.3 millimeter observations by combining ALMA 12-m array
extended (C40-9) and more compact (C40-6) configurations, resulting in
baselines ranging from 19~meters to up to 13.9~kilometers and a total
of 40-46 antennas. The combined observations are sensitive to spatial
scales of up 2\farcs0, with a spatial resolution of $\sim$20~mas. The
long baseline observations were acquired between Sept. and Nov. 2017
(Cycle 4) in four different blocks of $\sim$40~min each. Precipitable
water vapor ranged between 0.7 and 1.8~mm.  Observations of a phase
calibrator (J1826-2924) were alternated with the science to calibrate
the time-dependent variation of the complex gains. The cycling time
for phase calibration was set to 8~minutes and 54~seconds for the
compact and extended configurations, respectively. The ALMA correlator
was configured in Frequency Division Mode (FDM). Two spectral windows
with 1.875~GHz bandwidth were set up for detecting the dust continuum,
centred at 232.0~GHz and 218.0~GHz, respectively.

All data were calibrated by the ALMA staff using the ALMA Pipeline
version 40896 in the CASA package \citep{McMullin2007}, including
offline Water Vapor Radiometer (WVR) calibration, system temperature
correction, as well as bandpass, phase, and amplitude
calibrations. The short baseline and long baseline datasets were
calibrated independently. Self-calibration of the data was performed
to improve coherence (a single iteration of phase-only), after which
they were combined using the CASA task {\sc concat}. A positional
offset of (-20,-40) mas between the short and long baseline resulting
images was measured and corrected prior to combining the
datasets. This offset is consistent with the (-2.8,-38.0)~mas/yr
proper motion listed for this star (SIMBAD) and with the expected
astrometric accuracy of ALMA (typically a 10th of the synthesized
beam).

\subsection{ALMA continuum imaging}

Image reconstruction was initially performed using the CLEAN algorithm
(CASA version 5.2, task {\tt tclean}).  As the source is relatively
bright, we super-resolve the continuum data using our Maximum Entropy
Method (MEM) package {\sc uvmem}, which is part of the family of
algorithms based on maximum-entropy regularization. Here we used the
publicly-available GPU adaptation \textsc{gpuvmem}\footnote{The {\sc
    gpuvmem} code is publicly available and can be found here: {\tt
    https://github.com/miguelcarcamov/gpuvmem}.}
\citep[][]{Carcamo2018}, and regularized by minimizing the Laplacian
of the model image with an objective function.  The MEM reconstruction
provides the highest resolution without compromising on
sensitivity. We adopted the MEM image for our analysis. The final
image has a peak of 67~mJy~beam$^{-1}$ and an rms ($1\sigma$) of
14.7$\mu$Jy~beam$^{-1}$ for a synthesized beam of 27$\times$20~mas.
The total flux density over the whole image is 178$\pm$18~mJy (10\%
flux calibration uncertainty). The flux density of the central source
at 1.3~millimeter is 194$\pm$20\,$\mu$Jy. The bright ring at
$\sim$0\farcs22 has an integrated flux of 59$\pm$6~mJy, and the flux
density of the outer region is 126$\pm$13~mJy.  A side-by-side
comparison of the MEM versus CLEAN image reconstructions is shown in
Fig.\,\ref{supp:clean} (Appendix~\ref{app:clean}).

\begin{figure*}
  \centering\includegraphics[height=0.44\textwidth]{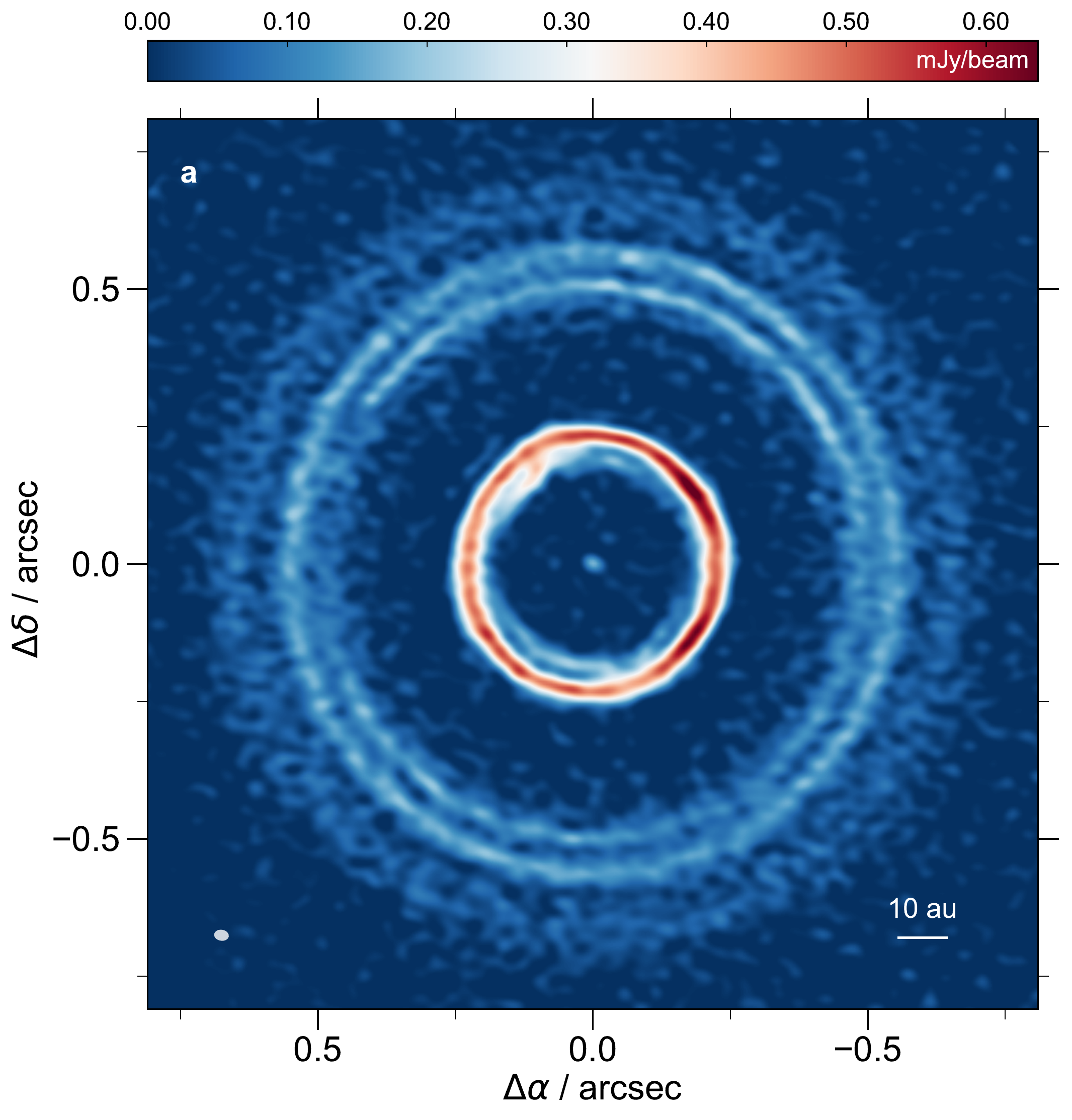}
  \includegraphics[width=0.49\textwidth]{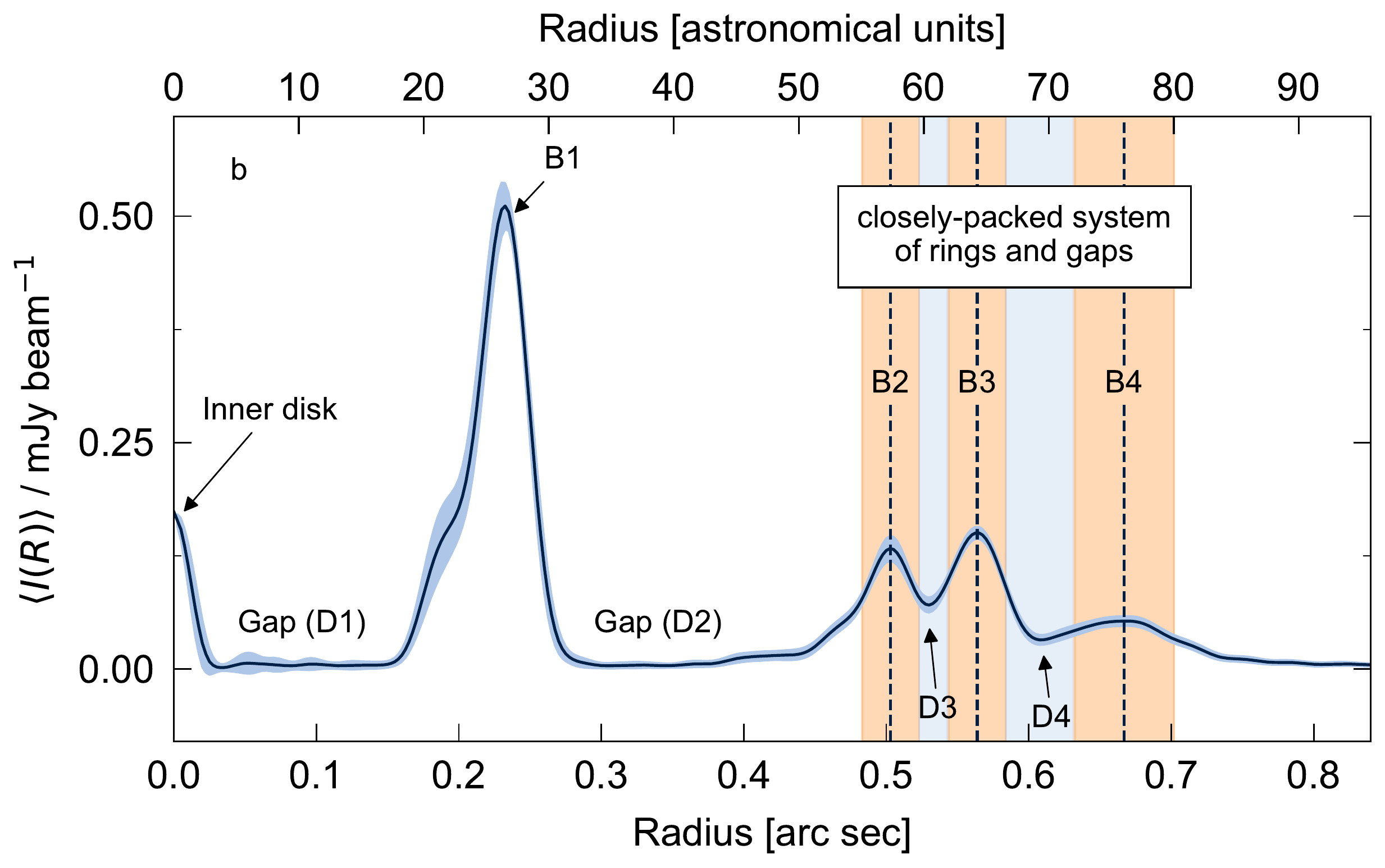}
  \caption{a.~Triple ring system in the outer region of the gapped
    disk around HD\,169142.  The 1.3~millimeter ALMA image is shown
    with a linear color scale in flux density units of mJy per beam,
    with a synthesize beam of 27$\times$20 mas (shown as an inclined
    ellipse in the bottom left corner). b.~Azimuthally averaged
    surface brightness profile. The shaded area around the profile
    corresponds to $\pm$3$\sigma$ dispersion around the mean. Rings
    and gaps are labelled. Unresolved continuum emission is detected
    at the expected location of the star (which reflects the
    synthesized beam or resolution). The vertical dashed lines show
    the locations of the fine outer rings. Shaded rectangles centered
    on B2, B3, and B4 show the full-width at half-maximum of each
    rings. }
    \label{fig:obs}
\end{figure*}

\subsection{A triple ring system in the disk's outer region}
\label{sec:res}

The ALMA image (Fig.\,\ref{fig:obs}a) reveals a prominent bright ring
(B1, see labelling in Fig.\,\ref{fig:obs}b) followed by substructure
in the outer region in the form of a system of three fine rings.  In
previous observations the three rings (B2, B3, and B4) were seen as a
single structure (we refer to this structure as the `outer region'),
with evidence of ripples at the last scattering surface of
micron-sized grains \citep{Pohl2017, Gratton2019}. In these new
observations, we see that the outer region is a packed system of three
rings, with mutual separations of $\sim$7~{\sc au} between B2 and B3,
and $\sim$12~{\sc au} between B3 and B4, for a distance to HD\,169142
of 113.9$\pm$0.8~pc \citep{Gaia2018}.

Rings B2 and B3 are detected at a 5 to 20\,$\sigma$ level
(1$\sigma$\,=\,15$\mu$Jy\,beam$^{-1}$), while emission from B4 reaches
4-5$\sigma$ over the image.  For a quantitative characterization of
the rings, we deproject the image by the disk inclination angle
$i$\,=\,$12.5^\circ$$\pm$0.5$^\circ$ and transform to polar
coordinates centred on the compact millimeter emission detected at the
location of the star (see Appendix~\ref{app:depro}). The polar
deprojections zoomed on each ring are shown in Fig.\,\ref{fig:depro}.
The disk azimuthally-averaged surface brightness profile is shown in
Fig.\,\ref{fig:obs}b.

The locations and widths of the fine rings are determined from
Gaussian fits to the average radial profile.  Three Gaussian
components were fit simultaneously plus a common low-order polynomial
to account for low level background emission.  Thus, the observed
locations of B2, B3 and B4 are 0\farcs503$\pm$0\farcs005 (57.3~{\sc
  au}), 0\farcs563$\pm$0\farcs005 (64.2~{\sc au}), and
0\farcs667$\pm$0\farcs009 (76.0~{\sc au}), respectively. Uncertainties
only represent the error in fitting the Gaussian centroids. Since the
rings are not circular (see below), the radii of each ring depend on
azimuth, varying within $\pm$0\farcs01 around the mean locations
quoted here.

The deconvolved width of each ring is determined by subtracting the
beam width $\sigma_{\rm beam} = {\rm FWHM}/\sqrt{8\log{2}}$ from the
best fit Gaussian width in quadrature.  Rings B2 and B3 have
deconvolved widths of only 1.5 and 1.8~{\sc au}, respectively, while
B4 is more radially extended with a deconvolved width of 3.4~{\sc
  au}. Rings B2 and B3 in HD\,169142 are likely some of the narrowest
structures in protoplanetary disks reported so far \citep[see DSHARP
  ring characterization in][where the narrowest ring is $\sim$3.4~{\sc
    au} in deconvolved width]{Dullemond2018}.

Fig.\,\ref{fig:depro} shows evidence for azimuthal dips along ring B2
spanning azimuthal angles $\sim$50$^\circ$ to 100$^\circ$ and
$\sim$180$^\circ$ to 270$^\circ$, measured counter clockwise from the
disk PA.  Interestingly, all three outer rings have a finite
eccentricity (0.09$\pm$0.03) and share a common focus. This is
demonstrated in Fig.\,\ref{fig:depro} which shows ellipses fit to the
rings' radii as a function of azimuthal angle (see
Appendix~\ref{app:depro} for details).

\subsection{A structured ring and two deep wide gaps in the disk's inner region}
\label{sec:B1}

\begin{figure*}[t]
  \centering \includegraphics[width=0.4\textwidth]{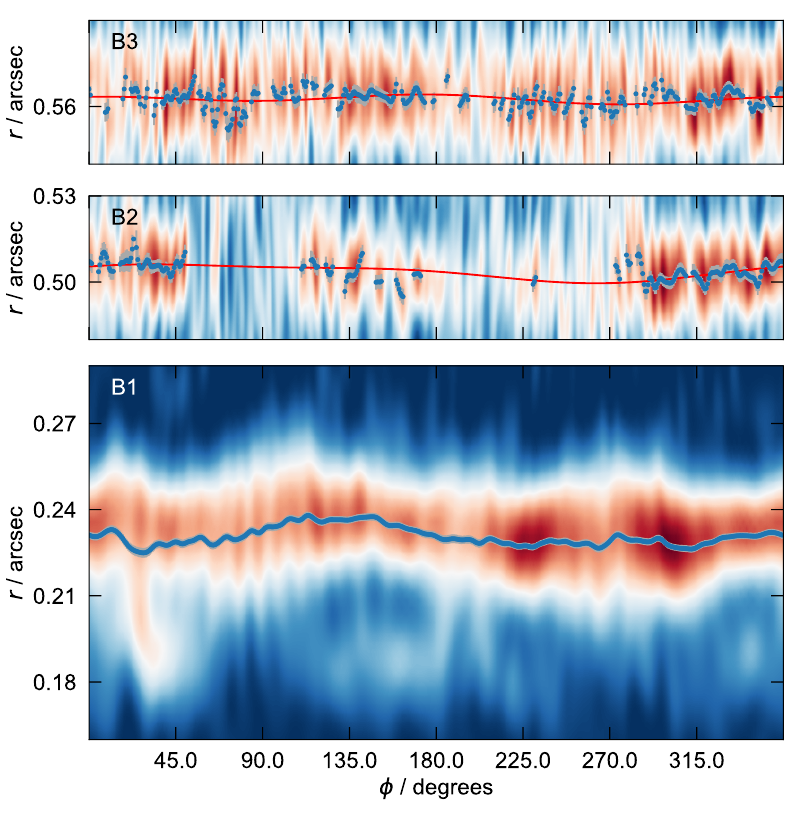}
  \hspace*{1.0cm}\includegraphics[width=0.35\textwidth]{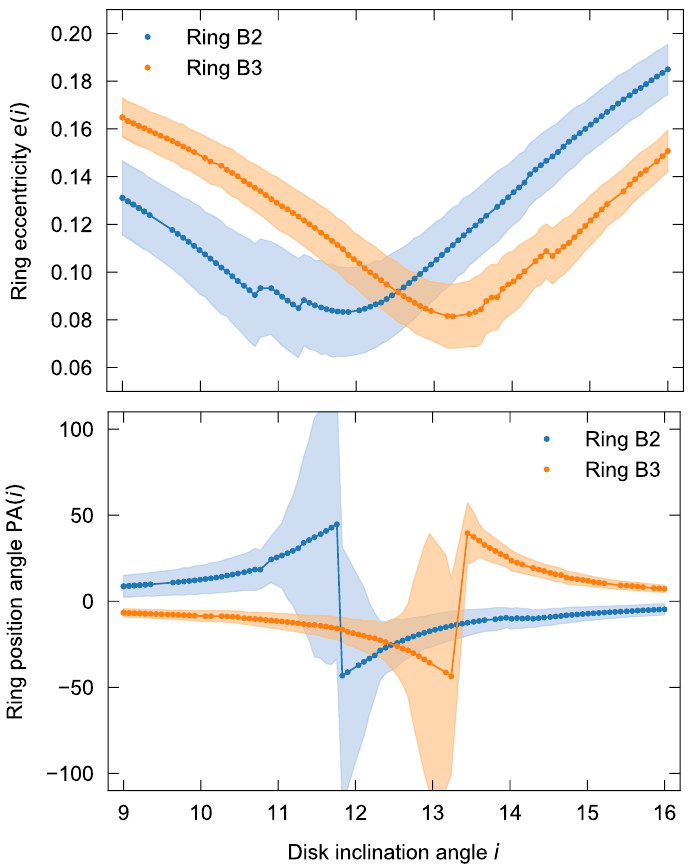}
  \caption{{\em Left.} Polar deprojection zoomed on rings B1, B2 and
    B3. The structure in B1 cannot be accounted for by neither a
    circular nor elliptical function as it is most likely dynamically
    perturbed by a giant planet. The ellipses that best fit the radial
    centroids of B2 and B3 have both $e=0.09$ and are shown as red
    solid lines.  Zero azimuth angle corresponds to position angle
    equal to 5$^\circ$ in Fig.\,\ref{fig:obs}, and increases east of
    north. Color scale is linear in mJy per beam, normalized to the
    peak of each ring for better visualization.  As ring B4 is not
    always detected above a 5 sigma level it is not shown here.  {\em
      Right.}  Intrinsic eccentricities (top) and position angles (PA,
    bottom) of rings B2 and B3 as a function of disk inclination
    angle. The shaded areas show $\pm2\sigma$ uncertainties. In the
    bottom panel, the origin of position angles is set to the disk PA
    (of 5$^\circ$ on the sky). Assuming B2 and B3 are co-planar, our
    fit constrains the inclination angle of the disk to
    $12.5\pm0.5^\circ$ ($1\sigma$ uncertainty). This inclination value
    is used in the deprojection shown in the left panel. At high and
    low inclinations, the ellipse PA approaches the disk PA, as
    expected.  }
  \label{fig:depro}
\end{figure*}

The brightest feature at 1.3 millimeter is the previously known inner
ring B1, whose remarkable radial and azimuthal structure is further
emphasized here at finer angular resolutions (Fig.\,\ref{fig:depro}).
Previous hydrodynamical modelling have suggested that this structure
is consistent with dynamical interactions with forming giant planets
located inside the gaps D1 and D2 \citep{Bertrang2018}. The narrowness
of ring B1 is explained by efficient radial trapping of dust in a
pressure maximum, which puts an abrupt halt to the inward radial drift
of dust particles $\geq$1\,mm \citep{Pinilla2012}. Interestingly, as
seen in Figs.\,\ref{fig:obs}a~and~\ref{fig:depro}, the structure in B1
may well be interpreted as a multiple ring. B1 is not split in the
same way as the outer ring, rather it shows a very irregular and faint
inner ring.

In addition, compact continuum emission is detected at the star's
location.  We interpret this emission as coming from thermal emission
from an inner disk of $\leq$1\,{\sc au} in radius, compatible with its
near-infrared excess \citep{Chen2018} as free-free emission is not
expected to be significant in HD\,169142 \citep{Osorio2014}.

The massive protoplanets required to explain the perturbed morphology
of B1 are unlikely to split the outer dust disk into the observed
triple ring structure \citep{Bae2018}. Instead, their dynamical
interaction with the disk should lead to the clearing of deep and wide
gaps such as D1 and D2, which are both $\sim$20~{\sc au} in width and
heavily depleted of dust {\em and} gas \citep{Pohl2017, Fedele2017,
  Ligi2018, Bertrang2018}.  Yet, with the fine angular resolution
provided by these new ALMA observations, we find that the outer region
of HD\,169142 is reminiscent of the ring systems in TW~Hya
\citep{Andrews2016} and HL~Tau \citep{ALMA2015}, both of which lack
the presence of a deep wide gap in the dust continuum.

\section{Hydrodynamical simulations and radiative transfer calculations} \label{sec:model}

\subsection{Theoretical background} \label{sec:tb}

Several models have been proposed to explain disks with
rings. Enhanced dust growth at condensation fronts can produce fine
rings \citep{Banzatti2015, Zhang2015, Okuzumi2016}, however, the
location of rings and gaps are uncorrelated with the expected
locations of snowlines \citep{Long2018, Huang2018, vandermarel2019},
and these rings are unlikely to be eccentric.  Sharp variations in the
gas viscosity associated with dead-zones \citep{Flock2015,Miranda2017}
and radially variable magnetic disk winds \citep{Suriano2018} also
yield ring-like structures, but they are driven by MHD instabilities
that occur mostly within the central $\sim$20\,{\sc au}.

Theory of disk-planet interactions predicts that a low-mass planet can
produce multiple dust rings by the propagation and dissipation of
spiral density waves \citep{Goodman2001,Dong2017,Dong2018}. The
dissipation of waves progressively moves dust away from the planet's
orbit, with the consequence that two narrow dust gaps form, one on
each side of the planet's orbit. The expelled dust thus forms two
rings exterior to the gaps. Also, dust can remain at the planet’s
(co-)orbital radius, forming a third ring between the two dust gaps
\citep{Dong2017}. Previous hydrodynamic simulations have explored this
scenario and found that, in a low-viscosity disk ($\alpha$ viscosity
of 10$^{-4}$ or lower), a single low-mass planet indeed produces
multiple rings in the dust distribution separated by annular gaps
\citep{Dong2017, Dong2018, Bae2018}, observable at ALMA
resolutions. The resulting ring system depends mainly on disk
viscosity, disk temperature (aspect ratio), planet mass, and the
elapsed time, which will dictate the number of additional spiral arms
and hence the potential number of gaps \citep{Bae2018}.

Given the sharpness and small widths of the rings in HD\,169142 (the
deconvolved width of ring B2 is only $\sim$1.5~{\sc au}), the $\alpha$
turbulent viscosity should be fairly small allowing a single low-mass
planet to produce multiple rings and gaps.  Assuming standard radial
profiles for the disk structure, a relation that connects the distance
between the two dust gaps, relative to the orbit of the planet, and
the planet's mass and disk scale-height can be found. Informed by
hydrodynamic calculations \citep[assuming $\alpha \la
  10^{-4}$,][]{Dong2018}, the relation reads
$$\frac{r_{\rm OG} - r_{\rm IG}}{r_{\rm p}} \approx 2.9 \left (
\frac{\gamma + 1 }{12/5} \frac{M_{\rm p}}{M_{\rm th}}\right)^{-2/5}
\left( \frac{h}{r} \right), $$
\noindent where $r_{\rm IG}$ and $r_{\rm OG}$ are the locations of the
inner and outer gaps around the planet, respectively, and $r_{\rm p}$
is the planet's orbit.  $M_{\rm th}=M_\star(h/r)^3$ is the disk
thermal mass, $\gamma$ is the polytropic index (equal to 1 for
isothermal gas), and $h/r$ is the disk aspect ratio.
Fig.\,\ref{fig:obs}b shows that the narrow gaps D3 and D4 are
separated by $\sim$9~{\sc au}. In the outer region, the observed flux
density of the rings translates into a brightness temperature of
8-12~K, consistent with a disk aspect ratio of 3\%.  Thus, for
HD\,169142's stellar mass ($M_\star\!\approx\!1.7\,M_\odot$), disk
aspect ratio ($h/r\!\approx\!0.03$), and approximating $r_{\rm p}$ by
B3's location at $\sim$64~{\sc au}, a planet of $\la$10\,$M_\oplus$
could produce gaps separated by $r_{\rm D4} - r_{\rm
  D3}\approx\,9$\,{\sc au}.

\subsection{Physical model and numerical setup}

We perform hydrodynamic simulations of an outer disk composed of gas
and a distribution of dust particles, to test whether the three bright
rings (B2, B3, B4) unveiled by our ALMA observations can be explained
by a single forming planet. The main observables are the distance
between the gaps, the mutual separation between the rings, and the
observed fluxes.  In this study, we do not attempt to model the
previously known inner features (D1 and B1) but rather focus on the
new substructure of the outer region.

\begin{figure*}
  \centering
  \includegraphics[width=.7\textwidth]{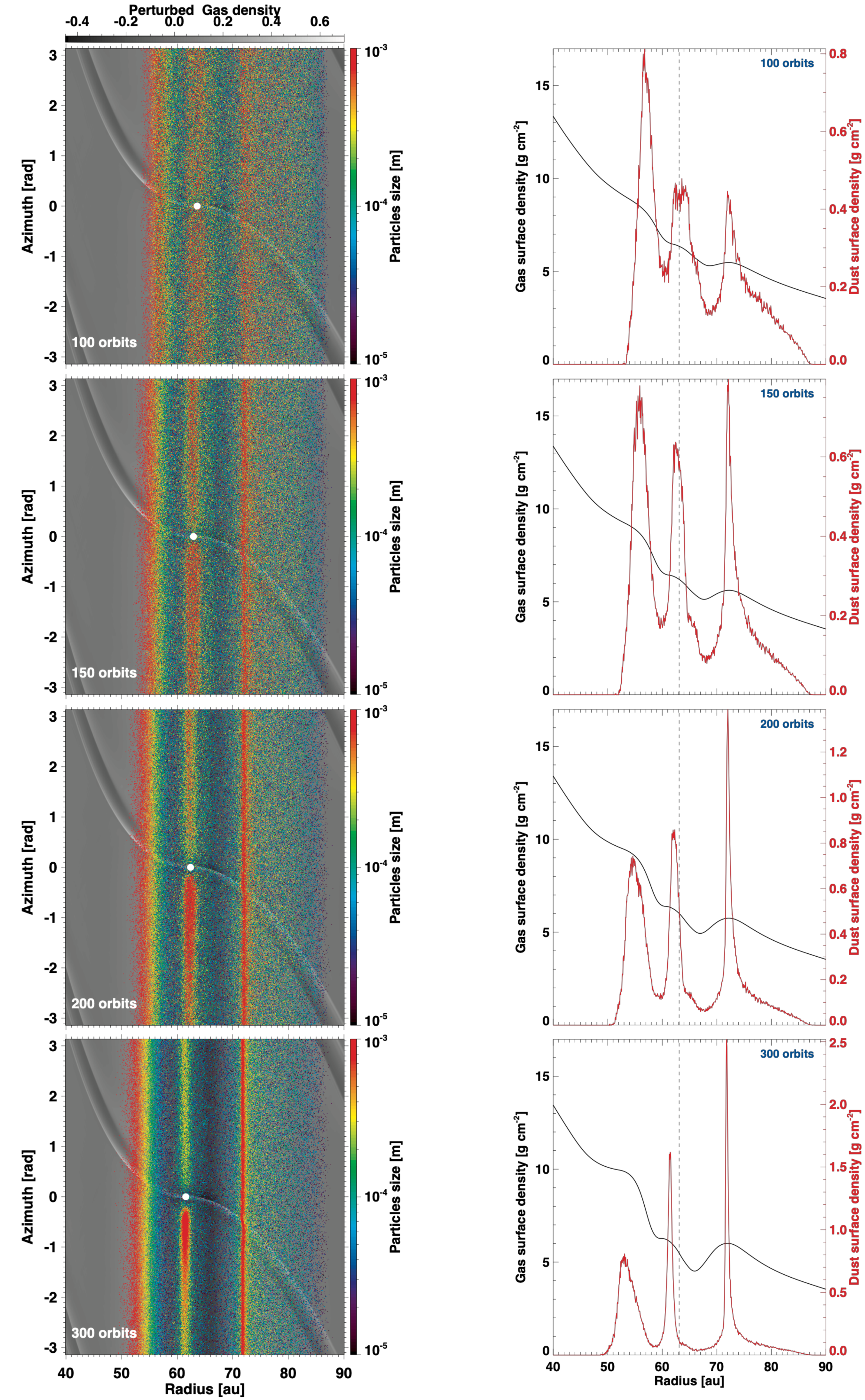}
  \caption{Gas and dust spatial distribution of a hydrodynamic
    simulation with a migrating 10$M_\oplus$ (mini Neptune size)
    planet. The perturbed gas surface densities relative to the
    initial profile are shown after 100 (a), 150 (b), 200 (c) and 300
    (d) orbits of the mini Neptune. Dust particles overlaid using
    colored dots (color varies with particle size, see color bar on
    the r.h.s. of the panel). The planet is marked by a white circle
    and can be spot through its spiral density waves.  Three dust
    rings are produced after 100 orbits: one at the planet's orbital
    radius ($\sim$64 au), and two on each side of the planet's
    gap. The two dust rings on both sides of the planet are not
    symmetrical w.r.t the planet's orbital radius. The asymmetry
    increases with time as the planet moves inward (the outermost ring
    keeps a fairly constant location).}
  \label{fig:hydro}
\end{figure*}

\subsubsection{Hydrodynamical simulations}

Our 2D hydrodynamical simulation of the gas and dust in the HD\,169142
disk has been carried out with the code Dusty FARGO-ADSG. It is an
extended version of the publicly available code FARGO-ADSG, which
solves the gas hydrodynamics equations on a polar mesh
\citep{Masset2000, Baruteau2008a, Baruteau2008b}, and which models
dust as Lagrangian test particles\footnote{The FARGO-ADSG code is
  publicly available at the address {\tt
    http://fargo.in2p3.fr/-FARGO-ADSG-}. The version including a
  Lagrangian treatment of the dust particles, Dusty FARGO-ADSG, can be
  made available upon request to co-author Clement Baruteau.}
\citep{Baruteau2016, Fuente2017}.  The simulation has been designed to
test whether the three bright rings (B2, B3, B4) unveiled by our ALMA
observations can be explained by a single forming planet. In our
simulation the star's mass is assumed to be 2~$M_\odot$.  We perform a
set of simulations sampling the main parameter space \{planet mass,
disk scale-height, disk alpha viscosity\}, although no exhaustive
exploration is intended as finding a perfect fit is not in the scope
of this work.

The gas momentum and continuity equations are solved on a polar mesh
with 900 cells logarithmically spaced between 35 and 105~{\sc au} in
radius, and 1,200 cells evenly spaced between 0 and 2$\pi$ in
azimuth. Non-reflecting boundary conditions are used to avoid
reflections of the planet wakes at the radial edges of the
computational grid.

The disk has an initial surface density profile decreasing with radius
$r$ as $r^{-1}$ with an exponential cutoff beyond 100~{\sc au}. The
initial surface density is 2.9~g~cm$^{-2}$ at 100~{\sc au}, very close
to the radiative transfer model which fits the CO istotopologue and
continuum observations at 1.3~mm \citep{Fedele2017}. This large
surface density means that Type I migration cannot be discarded,
especially for the low planet masses we are exploring. A locally
isothermal equation of state is used with the disk temperature
decreasing with radius as $r^{-1/2}$, and equal to 8~K at 70~{\sc au}
(motivated by the observed brightness temperature), slightly lower
than previous models. This low temperature, which is required to match
the small mutual separations between the bright rings, translates into
a disk aspect ratio profile in $r^{1/4}$ and equal to 0.033 at 69~{\sc
  au}. Gas self-gravity is included since the disk's Toomre
$Q$-parameter varies from $\sim$6 to $\sim$9 throughout the
computational grid. The $\alpha$ turbulent viscosity is set to
$10^{-5}$.  This low viscosity is motivated by the sharpness of the
observed rings. Accretion onto the planet is not included.

The code solves the equations of motion for 200,000 dust particles
with radii between 10~$\mu$m and 3~mm. Dust particles feel the gravity
of the star, the planet, the disk gas (since gas self-gravity is
included) and gas drag. Dust turbulent diffusion is also included as
stochastic kicks on the particles position vector
\citep{Charnoz2011}. However, dust self-gravity, dust drag, growth and
fragmentation are not taken into account. Dust particles (assumed to
be compact spheres of 2~g~cm$^{-3}$ internal density) are inserted at
the beginning of the simulation with a number density profile
decreasing as $r^{-2}$ between 59 and 90~{\sc au}. This corresponds to
an initial dust surface density decreasing as $r^{-4}$.  Although
quite steep, the surface density evolves into a profile which
increases with $r$ between the rings after 150 orbits. See the
perturbed gas density as well as the distribution of dust particles
shown in Fig.\,\ref{fig:hydro}. Dust feedback onto the gas is
discarded as the dust's surface density along the B2, B3, and B4 rings
remains comfortably smaller than the gas surface density in our model
(see Fig.\,\ref{fig:hydro}), for the gas-to-dust mass ratio assumed in
the radiative transfer calculation (see Section~\ref{sec:rt}).

A low-mass planet of a mini-Neptune mass (motivated by
Sec.~\ref{sec:tb}) is inserted at 69~{\sc au} at the beginning of the
simulation. The planet-to-star mass ratio is set to
1.7$\times$10$^{-5}$, which for a 2\,$M_\odot$ star translates into
11~$M_{\oplus}$\footnote{The stellar mass of HD\,169142 is closer to
  $M_\star=1.7~M_\odot$, which yields a planet mass of
  9.6\,$M_{\oplus}$ in our simulation. Hence we round the
  mini-Neptune's mass off to $\sim$10\,$M_\oplus$ throughout the
  paper.}. The planet migrates due to disk-planet interactions and
reaches $\sim$64~{\sc au} after 150 orbits, which is very close to the
location of the B3 ring.  Higher mass planets (here we considered
planet-to-star mass ratio of 6$\times$10$^{-5}$, i.e., a
$\sim$33\,$M_\oplus$ planet) produce significantly wider gaps even at
higher viscosities (a simulation with $\alpha=10^{-4}$ was performed,
not shown here).

The time that best reproduces our observations of the outer rings is
approximately at 150 orbits of the mini Neptune, i.e.,
$\sim$0.05\,Myr. We note that the rings are present since orbit
$\sim$100 onwards (see Fig.~\ref{fig:hydro}).  The model is stable for
at least 1000 planet orbits ($\sim$0.4\,Myr). After this, a global
$m$=$1$ mode grows in the disk, leading both the planet and the disk
to develop some eccentricities. After this, the co-rotating ring
gradually diffuses with time as the gas surface density decreases.

\subsubsection{Radiative transfer calculation}\label{sec:rt}

We compute the continuum emission from our dust simulations using the
public radiative transfer (RT) code {\sc radmc3d} (version
0.41). Twenty logarithmically-spaced bins are used to allocate the
200,000 dust particles from the simulation. The dust sizes ($s$)
follow a power-law distribution $n(s)\propto s^{-3.5}$, with minimum
and maximum sizes of $1\,\mu{\rm m}$ and $1{\rm mm}$, respectively.
Assuming a gas-to-dust ratio of 33 yields a total dust mass in the
outer region of $\sim$100$M_\oplus$.

Each 2D distribution of dust particles is turned into a surface
density and expanded in the vertical direction assuming hydrostatic
equilibrium. The dust's scale height $h_{i,{\rm d}}$ is size-dependent
and follows
\begin{equation}
h_{i, {\rm d}} = h \times \sqrt{\frac{D_{\rm z}}{D_{\rm z} + {\rm St}_i}},
\label{eq:Hd}
\end{equation}
\noindent where $h$ is the initial gas pressure scale-height, ${\rm
  St}_i$ is the averaged Stokes number for the $i^{\rm th}$ dust size
bin, and $D_{\rm z}$ is a turbulent diffusion coefficient in the
vertical direction which depends on the level of turbulent activity
across the vertical extent of the disk \citep{Yang2018}. We assume
$D_z$ to be proportional to the $\alpha$ turbulent viscosity in our 2D
hydrodynamic simulation \citep[see][for more details]{Baruteau2019}. A
$D_z$ value equal to 10\,$\alpha$ yields dust temperatures which match
the observed brightness temperature of rings B2 and B3. In the RT
calculation, we use 24 cells logarithmically-spaced to sample 3
scale-heights in colatitude (with finer sampling towards the
midplane).

We expand the hydrodynamic simulation's grid to include the inner
regions and then artificially add the inner disk and the bright ring
B1. Their parameterisation follows previous RT modelling which fits
the 1.3 millimeter data \citep{Fedele2017}. The presence of these
structures that block stellar radiation are needed to recover the low
temperatures of the outer regions.  A 7320~K star with a radius of
1.5~$R_\odot$ \citep{Gaia2018} is placed at the centre of the grid and
$10^8$ photon packages are propagated to compute the temperature of
each dust bin size via the Monte Carlo {\sc radmc3d} {\tt mctherm}
task.  The temperatures range between $\sim$13 and $\sim$20~K in the
outer rings, consistent with previous multi-wavelength modelling
\citep{Fedele2017}. These values are 2$\times$ higher than the locally
isothermal prescription used in the hydrodynamical simulation, which
is representative of the gas scale-height, rather than a physical
temperature of dust grains.  Ray-tracing is performed to solve the
radiative transfer equation for continuum emission assuming the dust
is a mix of 30\% amorphous carbons and 70\% silicates. The resulting
synthetic image is convolved with the MEM resolution beam and it is
shown in Fig.\,3. The high fidelity of the ALMA image allows direct
comparison with the data in the image plane.  The hydrodynamical model
reproduces the locations of B2, B3, and B4 as well as the level of
flux along rings B2 and B3, and to a lesser extend ring B4's.  This
can be best appreciated in Fig.\,\ref{supp:depro}
(Appendix~\ref{app:giant}).

\subsection{Results} \label{sec:model2}

\begin{figure*}
  \centering
  \includegraphics[height=0.46\textwidth]{fig1a.pdf}
  \includegraphics[height=0.46\textwidth]{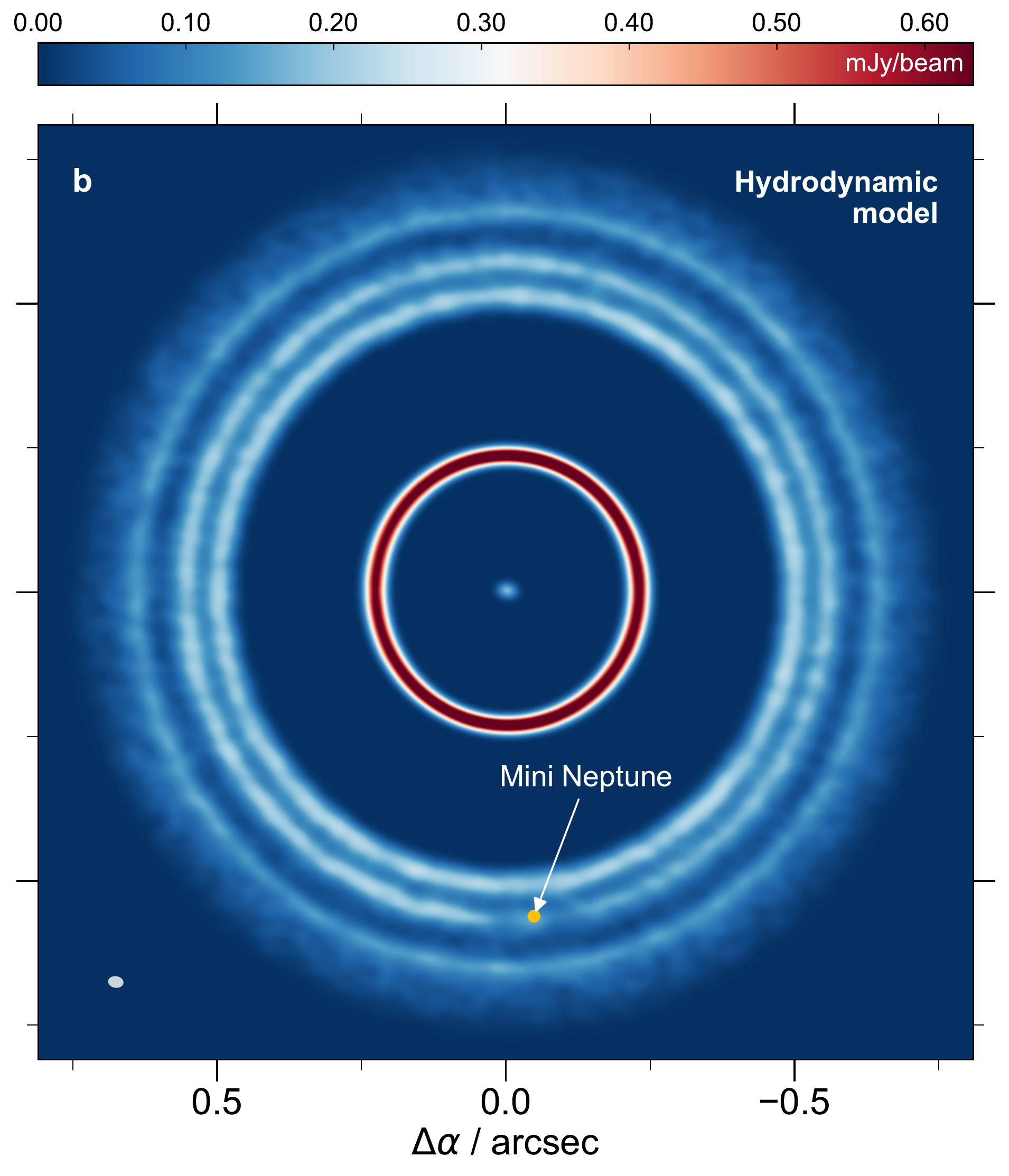}
  \caption{Side-by-side comparison between the ALMA 1.3 mm image of
    HD\,169142 ({\em left}, same as in Fig.~\ref{fig:obs}) and the
    predicted image based on the hydrodynamic simulation of a
    migrating mini-Neptune planet of 10~$M_\oplus$ ({\em right}).
    After 150 orbits, the triple ring system emerges in the outer
    disk. The inner disk and ring B1 are not included in the
    hydrodynamic simulation but added {\em a posteriori} at the
    radiative transfer stage.  }
\label{fig:model}
\end{figure*}

A hydrodynamical model including a 10~$M_\oplus$ planet embedded in a
disk with an aspect ratio $h=0.033$ reproduces the observations
(Fig.\,\ref{fig:model}).  The initial condition is such that the
planet is introduced at 69\,{\sc au}. After about fifty thousand
years, the planet migrates inwards to a $\sim$64~{\sc au} orbit,
shepherding ring B3 and ``pushing'' rings B2 and B4, thus explaining
the asymmetry in the rings mutual separations. This can be appreciated
in Fig.\,\ref{fig:hydro}. A simulation with a planet on a fixed orbit
(not included) produces rings that are too equidistant for the disk
and planet parameters assumed here.  If the 10\,$M_\oplus$ core formed
closer to the star and then migrated outwards, it would not be
consistent with the rings locations: the observed asymmetry reveals
the direction of migration.  This migration could have distinct
observational signatures in multiwavelength observations, for example,
by probing the spatial segregation of larger grains with longer
wavelength observations. Such predictions have recently been reported
by \citet{Meru2019} for a planet migrating in a high viscosity
disk. Despite the little knowledge we have about the initial
conditions of the dust density distribution, the model reproduces the
observed fluxes along B2 and B3, albeit with a slightly brighter ring
B4. The initial amount of large particles near B4's location is likely
overestimated in our simulations.

As the planet continues to interact dynamically with the dust
reservoir, it also starts clearing an {\em azimuthal} opening in its
vicinity. This produces a dip in azimuth along the middle ring at the
planet's location. Whether this dip is observable at 1.3~millimeters
depends on the planet's mass, the maximum size in the dust
distribution, and the elapsed time. The more massive the planet, the
faster and deeper is the opening of the azimuthal dip. On the other
hand, a large maximum particle size produces a more pronounced
dip. Dust growth and fragmentation, which are not taken into account
in our simulations, probably also have an effect on the observability
of this azimuthal opening. The lack of a clear azimuthal dip along
ring B3 indicates that the bulk of particles we observe have sizes
$<$1\,mm, the planet is rather small ($\la$10\,$M_\oplus$), or that it
may have formed less than a few hundred orbits ago ($\sim$50~kyr), or
most likely a combination of these options. Note that the
signal-to-noise ratio at the rings location might not be sufficient to
probe the azimuthal structure along B3. Longer wavelength observations
might be able to trace the location of the mini-Neptune as the larger
grains pile-up behind the planet (note mm particles in
Fig.\,\ref{fig:hydro}).

\subsection{Discussion}

\subsubsection{Could rings B2, B3 and B4 be shaped by two low-mass planets?}

A scenario where two low mass planets (one for each narrow gap)
results in the triple ringed structure is unlikely for several
reasons.  The small separation between the gaps (only 9~{\sc au})
requires scale-heights at the location of the planets which would be
too small for a realistic disk temperature, these translate into
$<$5~K at 64~{\sc au} \citep[following eq. 16 in][]{DongFung2017}. In
other words, if you put two planets into the two gaps, they will
likely open one common gap instead of two. Moreover, the similarity
between D3 and D4 suggests that if they are produced by two planets,
these ought to have formed at similar times and with similar
masses. The dynamical interactions between the closely orbiting
planets might also make them unstable.

\subsubsection{How does an inner giant planet impact the structure of rings B2, B3 and B4?}

In recent years, it has been shown that planet perturbations can
extend far beyond the Hill radius if the disk viscosity is low
\citep{Bae2018}. Since the model presented here assumes a very low
viscosity, any inner giant planet (in D2, for example) could have an
impact on the outer triple ring system. In a debris disk or a
protoplanetary disk with low gas surface density, giant planets could
potentially produce fine structures by mean motion resonances with the
dust. However, HD\,169142 is gas rich and the dust in the outer region
is still coupled to the gas distribution.  To further test the
possibility that the fine structure is somehow related to inner giant
planets, we included a giant planet in our simulation, of roughly a
Jupiter mass located in the middle of D2 at 38~{\sc au} (see
Appendix~\ref{app:giant}). The giant indeed opens a deep gap in the
gas, but fails to generate any additional narrow rings in the outer
disk. However, the spiral wakes excited by the giant and the vortices
at the gap edge (produced by the Rossby wave instability) induce
eccentricity on the outer triple rings associated with the
mini-Neptune (Fig.~\ref{supp:giant}). Recent scattered light imaging
indeed shows faint spiral features at the disk surface which can be
associated to putative giant planets in the wide gaps
\citep{Gratton2019}. The dust particles around the mini-Neptune can
acquire non-circular trajectories either (i) by direct gravitational
interaction with the mini-Neptune, whose eccentricity varies on
account of the inner giant's wakes depositing energy and angular
momentum in the planet's horseshoe region (Fig.~\ref{supp:orbit}),
and/or (ii) by being deflected by the (shock) wakes upon crossing
them.

The addition of a giant planet adds significant complexities as the
final configuration of rings would strongly depend on the initial
conditions for the dust distribution and the precise timing of the
formation of the mini Neptune and the giant planet.  As protoplanetary
disk observations grow in sensitivity and resolution, more and new
ingredients need to be considered in the modeling efforts.  This paper
is meant as a proof of concept to show that a simple one-planet
simulation can account for seemingly complex sub-structure in the
outer ring of a transition disk. Finding a perfect fit to the data is
not in the scope of this work. Indeed, some important aspect have been
left aside such as: dust growth \citep[e.g.,][]{Bae2018} and
fragmentation, dust back-reaction onto the gas
\citep[e.g.,][]{Gonzalez2017, Dipierro2018}, dust torque due to
scattered pebbles \citep{Benitez2018}, 3-D and MHD effects
\citep[e.g.,][]{Flock2015, Miranda2017}, and possibly many more. The
inclusion of planet migration was needed here to explain the asymmetry
in the rings' mutual separations, a step forward towards understanding
how low-mass planet migration can be studied from observations of dust
radial structures. However, a proper account of dust dynamics, planet
accretion and thermodynamics, in a self-consistent way, is necessary
to build the migration history of these low-mass protoplanets
\citep{Benitez2018}.

\subsubsection{What can these observations tell us about the planet formation process?}

The low mass of the putative protoplanet in HD\,169142, in addition to
the lack of clear evidence for structures associated with
gravitational instabilities in (abundant) multi-wavelength
observations of this source, suggests that a bottom-up process such as
core accretion could be responsible for its formation. This implies
that core accretion can potentially operate at $\sim$65~{\sc au}
within the age of the system \citep[$\sim$6~Myr,][]{Grady2007},
possibly assisted by pebble accretion as the outer region banks more
than 100\,$M_\oplus$ worth of pebble-sized solids (subject to
uncertainties of the dust opacities). The modeling presented here also
shows that the mini-Neptune should have formed well after the giant
planets carved gaps D1 and D2.  The presence of a giant planet in D2
would have produced a pressure maxima in the outer region which could
have enhanced core accretion via a dust trap \citep{Pinilla2012}.

\section{Concluding remarks} \label{sec:summary}

The new ALMA observations presented here show that a transition disk
with wide deep gaps can also host narrow-ring structures in its outer
region, similar to those observed in the HL~Tau and the TW~Hya
systems.

The HD~169142 observation allows to link the architecture of
protoplanetary disks with low mass planets. The interpretation via
hydrodynamics is a proof of concept that links the structure of
closely packed double gaps and tripple rings with a single and
migrating low-mass planet. Planetary migration naturally explains the
distinct mutual separations between the narrow rings.  The connection
was made possible in HD\,169142 thanks to the isolation of its outer
region. In the absence of a clear gap that separates an outer ring,
the superposition of multiple rings due to several planets hampers
simple and clear explanations such as that found for HD\,169142.

In HD\,169142, we have thus found evidence that suggests that low-mass
planet formation can occur in the outer regions of disks bearing
evidence for giant protoplanets. The planet formation mechanism,
likely core accretion or any bottom-up process, can thus produce
planet embryos at $\sim$65\,{\sc au} and outside the orbit of inner
giants, at least in certain disks.

\acknowledgments

We thank Ed Fomalont, Anya Yermakova and Philipp Weber for useful
discussions, as well as our anonymous referee for their constructive
comments. Financial support was provided by the government of Chile
grants Millennium Scientific Initiative RC130007, CONICYT-Gemini
32130007, and CONICYT-FONDECYT grant numbers 1171624, 1171246 and
1191934. S.P acknowledges support from the Joint Committee of ESO and
the Government of Chile. The data analysis and some of the simulations
were carried out in the Brelka cluster, hosted at DAS/U. de Chile
(Fondequip EQM140101).  Numerical simulations were carried out on the
CalMip machine of the Centre Interuniversitaire de Calcul de Toulouse,
which is gratefully acknowledged. This work was performed in part at
the Aspen Center for Physics, which is supported by National Science
Foundation grant PHY-1607611. This paper makes use of the following
ALMA data: ADS/JAO.ALMA\#2016.1.00344.S. ALMA is a partnership of ESO
(representing its member states), NSF (USA) and NINS (Japan), together
with NRC (Canada), MOST and ASIAA (Taiwan), and KASI (Republic of
Korea), in cooperation with the Republic of Chile. The Joint ALMA
Observatory is operated by ESO, AUI/NRAO and NAOJ. The National Radio
Astronomy Observatory is a facility of the National Science Foundation
operated under cooperative agreement by Associated Universities, Inc.
This work has made use of data from the European Space Agency (ESA)
mission {\it Gaia}, processed by the {\it Gaia} Data Processing and
Analysis Consortium (DPAC). Funding for the DPAC has been provided by
national institutions, in particular the institutions participating in
the {\it Gaia} Multilateral Agreement.

\vspace{5mm}
\facilities{ALMA Observatory}

\software{GPUVMEM \citep{Carcamo2018}, CASA \citep{McMullin2007},
  FARGO-ADSG \citep{Baruteau2008a,Baruteau2008b}, RADMC-3D
  \citep{Dullemond2012}.}

%% Appendix material should be preceded with a single \appendix command.
%% There should be a \section command for each appendix. Mark appendix
%% subsections with the same markup you use in the main body of the paper.

%% Each Appendix (indicated with \section) will be lettered A, B, C, etc.
%% The equation counter will reset when it encounters the \appendix
%% command and will number appendix equations (A1), (A2), etc. The
%% Figure and Table counter will not reset.

\appendix

\section{Comparison between CLEAN and MEM algorithm} \label{app:clean}

\begin{figure}[ht!]
  \centering
  \includegraphics[height=.365\textwidth]{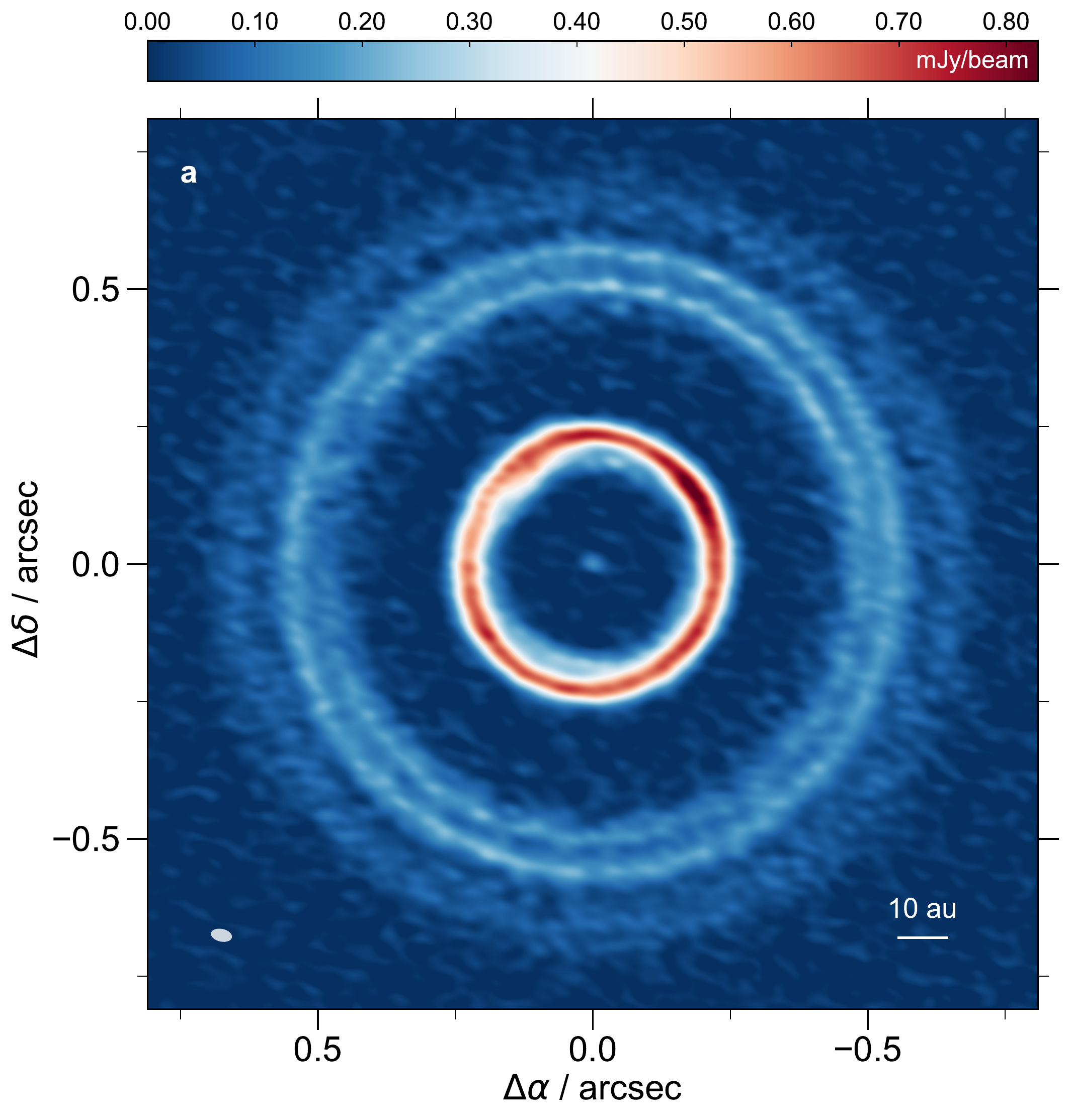}
  \includegraphics[height=.365\textwidth]{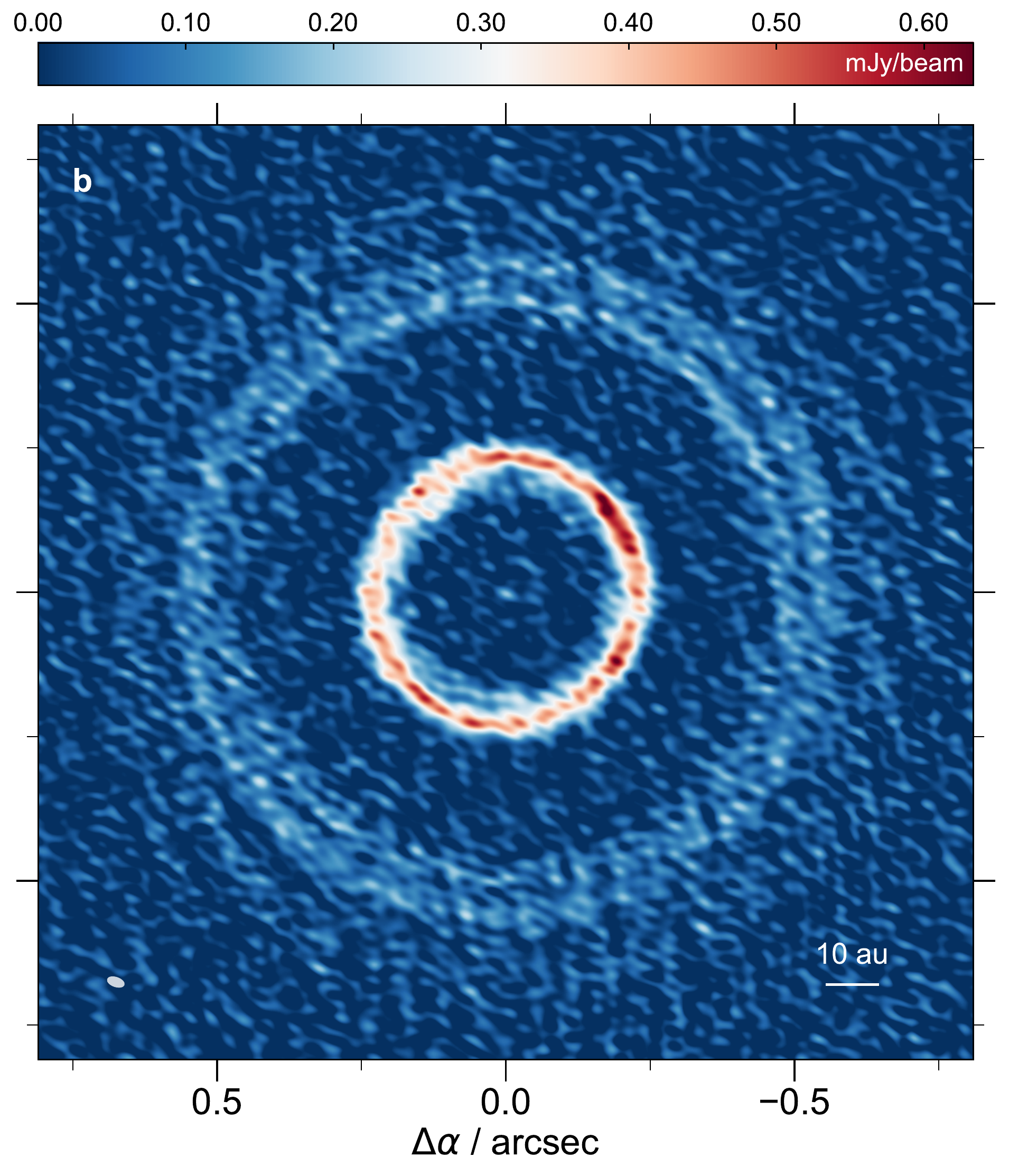}
  \includegraphics[height=.365\textwidth]{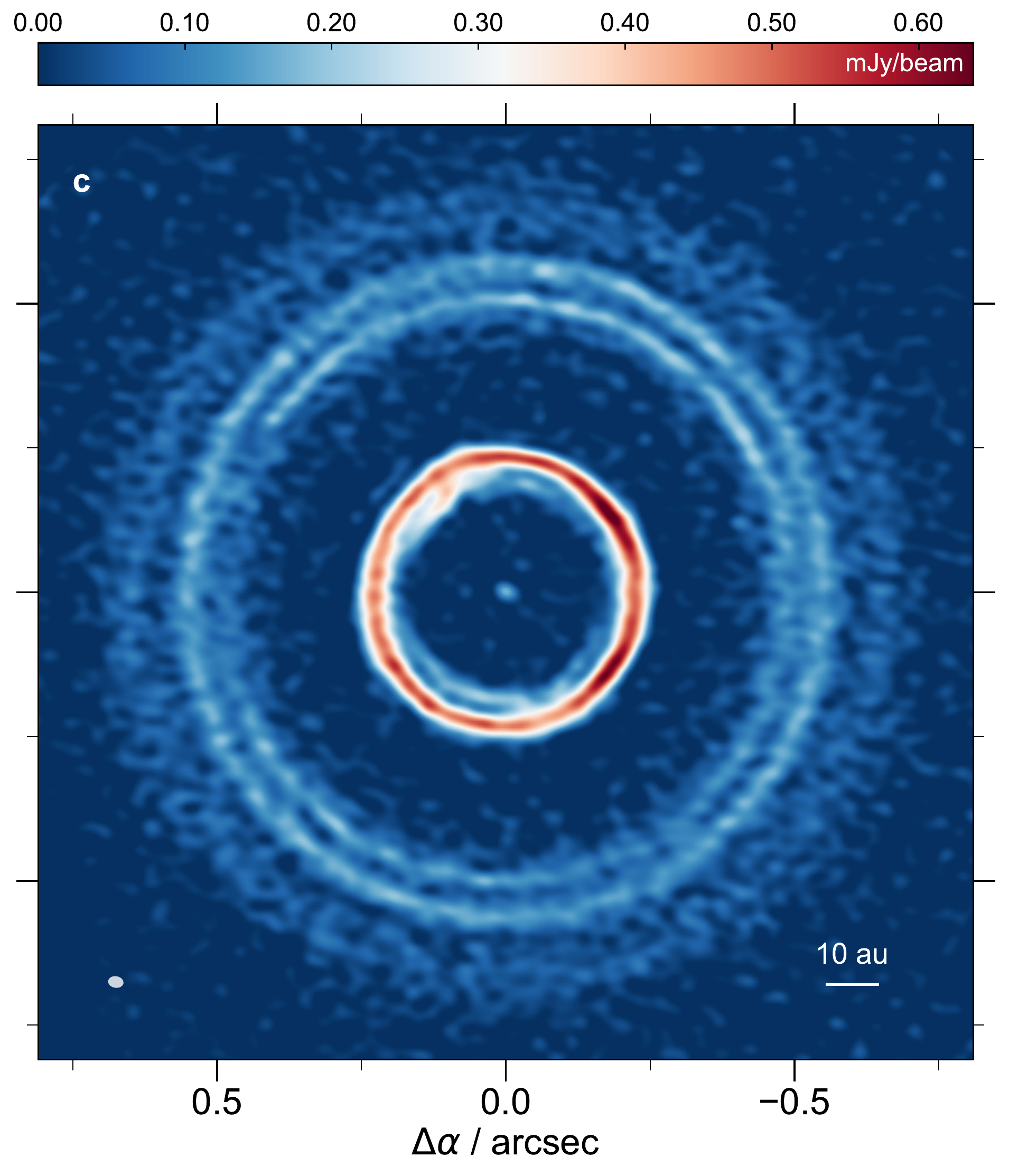}
  \caption{HD\,169142 image synthesis.  CASA {\tt tclean} with Briggs
    weighting, with robust parameter 0.0, yielding a resolution of
    39$\times$23~mas (a), uniform weight with robust -2.0 and
    resolution 32$\times$18~mas (b), and the {\sc gpuvmem}
    deconvolution (c). The effective resolution of the MEM deconvolved
    image is 27$\times$20~mas and is measured directly from a 2D
    Gaussian fit to the stellar component in the centre of the
    field. }
  \label{supp:clean}
\end{figure}

A side-by-side comparison of HD\,169142 image synthesis using CLEAN
and MEM algorithms is shown in Fig.~\ref{supp:clean}. The triple ring
structure in the outer region is robust in all image
reconstructions. The `uniform' weighting CLEAN image has a comparable
resolution to MEM but at the cost of decreasing sensitivity.

\section{Polar deprojection and ellipse fitting} \label{app:depro}

To study the eccentricity of the rings, we extract the radial profile
of each ring as a function of azimuth, $R(\theta)$, by fitting
gaussian functions in radius (using the {\sc Python} {\tt iminuit}
optimization library).  Only signal above $5\sigma$ is included in the
process. Ring B4 is not considered as it only has a small number of
points with high signal-to-noise. B1 is also left aside as its
perturbed morphology cannot be reproduced by a simple elliptical
curve. The optimization yields the values for $R(\theta)$ and their
error bars $\Delta R(\theta)$, for each ring.  The ellipse equation is
then fitted to $R(\theta)$ for a given ring using a direct least
square procedure \citep{Fitzgibbon1996}. The fitting procedure yields
the ellipse centre, its angle of rotation and its eccentricity
($e=\sqrt{1-(b/a)^2}$ where $a$ and $b$ are the semi-major and
semi-minor axes, respectively).  The best fit ellipse for each ring
are plotted on the polar projection in Fig.\,\ref{fig:depro}.  The
uncertainties in the eccentricity values are calculated using a
Monte-Carlo approach where data points are modified by random numbers
of the order of $3\Delta\theta$ and then fitted following the
aforementioned procedure. This is repeated a thousand times and the
final errors are drawn from the standard deviation. Correlated noise
is accounted for by multiplying the centroids' error by a factor
$\sqrt{N_{\rm pix}}$, where $N_{\rm pix}=5$ is the number of pixels
along the synthesized beam's major axis.

The uncertainty of the eccentricity is dominated by our imprecise
knowledge of the inclination angle of the disk. In order to determine
the significance of the fitted values, we explore a range of
inclination angles ranging between 9$^\circ$ and 16$^\circ$, repeating
the fitting procedure described above for each inclination value.  The
results are shown in Fig.\,\ref{fig:depro}b.  Rings B2 and B3 show
eccentricities which indeed vary with inclination angle, with a
minimum eccentricity of $e$$\approx$$0.085$. Interestingly, the rings
have different eccentricities for a given inclination, except at
$i$$\approx$$12.5^\circ$, where both rings have the same
eccentricity. This inclination angle is close to the standard value of
$i$$\sim$$13^\circ$ used in previous multi-wavelength modelling
\citep[e.g.][]{Fedele2017, Bertrang2018}. The intrinsic position
angles of the ellipses also coincides at
$i$$\approx$$12.5^\circ$. This fitting procedure thus suggests a
common inclination for both B2 and B3 of
$i$$\approx$$12.5^\circ\pm0.5\deg$ (1$\sigma$). At this inclination,
the eccentricity of B2 and B3 share the common value of 0.09$\pm$0.02
(2$\sigma$ uncertainty, same as the shaded area in
Fig.\,\ref{fig:depro}). A disk inclination angle different from
$12.5^\circ$ yields different eccentricity values for B2 and B3, which
would suggest the occurrence of warping in the outer regions.

\section{Addition of a giant planet associated with gap D2} \label{app:giant}

\begin{figure}
  \includegraphics[width=\textwidth]{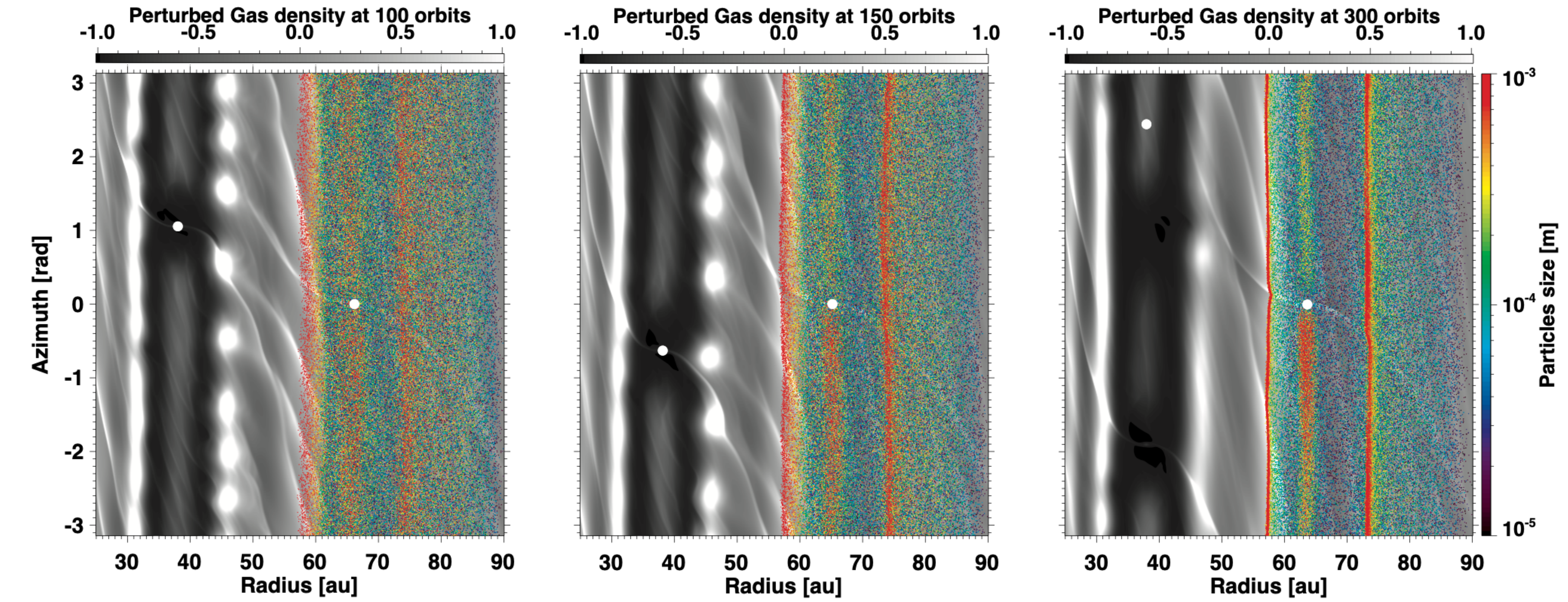}
  \caption{Effect of an inner giant planet on the triple ring
    structure over 100~kyr. The figure shows the perturbed gas density
    and the distribution of dust particles after 100 ({\em left}), 150
    ({\em middle}), and 300 ({\em right}) orbits of the
    mini-Neptune. The onset of the ring structure around the
    mini-Neptune happens near 100 orbits, while at 300 orbits
    (100~kyr) the ring B3 develops an azimuthal asymmetry, similar as
    in the case without the giant planet. In the presence of the
    giant, the outer rings associated with the mini-Neptune interact
    with the giant's wakes, acquiring some eccentricities. The gas
    distribution in the outer region also becomes eccentric.  }
  \label{supp:giant}
\end{figure}

The impact that an inner giant could have on the outer narrow rings is
explored by including a 0.5~$M_{\rm Jup}$ planet fixed at 38~{\sc au},
in the middle of the observed gap D2, in addition to the mini-Neptune.
The general setup inherits the same parameters as the single
mini-Neptune simulation, with some changes. The initial radius of the
orbit of the mini-Neptune planet is decreased to 68~{\sc au}, this is
to account for a slightly lower migration rate in the presence of the
inner giant.  The grid resolution is lowered to 600$\times$900 cells
in radius and azimuth, respectively. The grid's inner edge is extended
down to 20~{\sc au} as to include gap D2 in the radial domain.  The
smaller inner edge translates into a significant computational
cost. The number of particles is decreased by half (i.e only 100,000
particles are used) so as to make the computation less expensive.
Dust particles are distributed over the same region as in the previous
simulation.  Dust around the giant planet is not included as this
would require increasing the number of particles, making the
simulation too expensive. As mentioned in the main text, fitting the
already known features B1 and D1 is beyond the scope of this
work. Modeling all features in HD\,169142 system requires an extensive
search of a large parameter space, but the main difficulty lies on the
little knowledge available of the initial conditions and the relative
age of each planet. In our simple two-planet setup, both companions
are introduced simultaneously at the beginning of the simulation.

\begin{figure}
  \centering {\small \hspace*{-0.4\textwidth} {a.} \hspace*{0.4\textwidth} {b.}}\\
  \includegraphics[height=0.25\textheight]{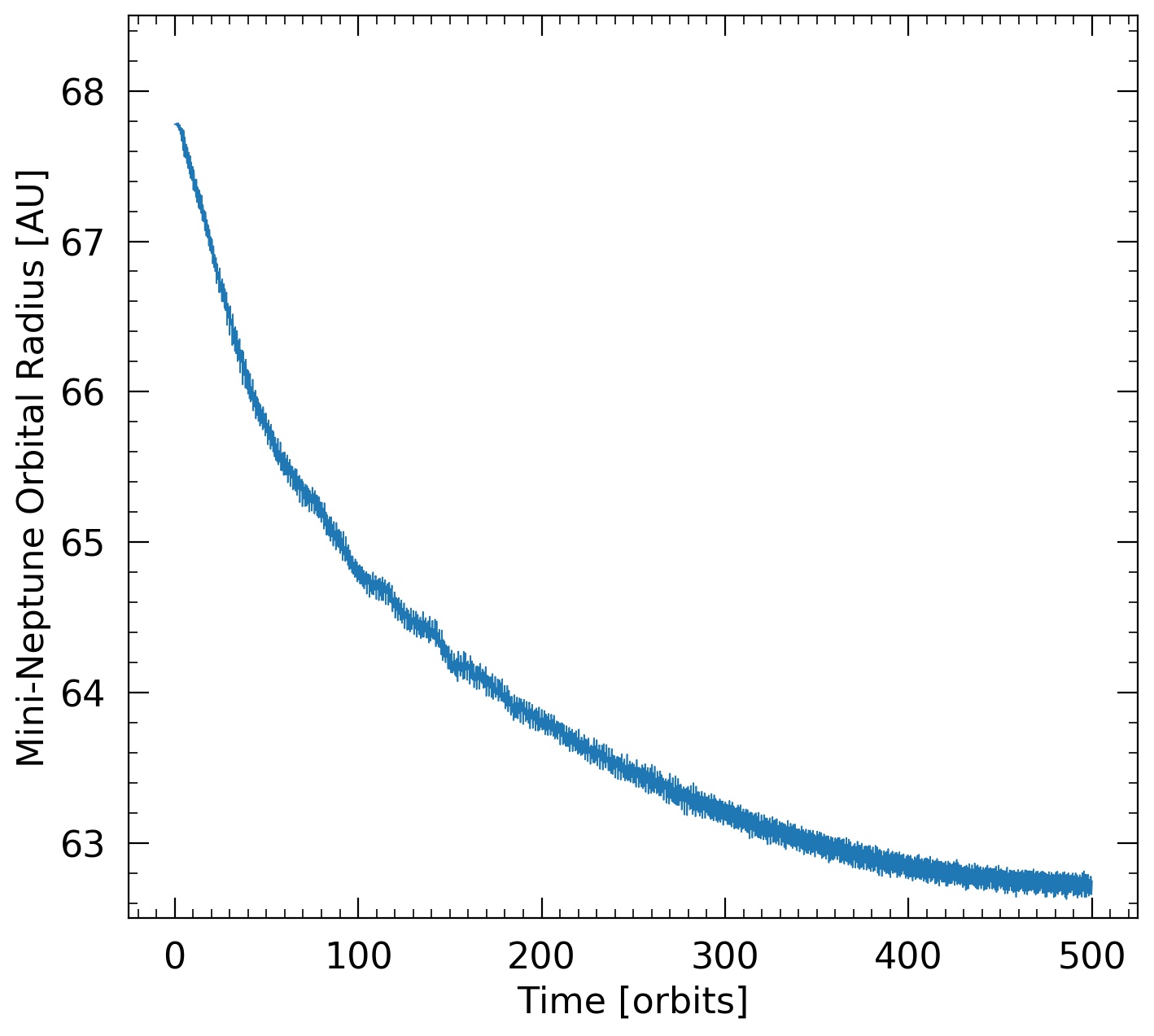}
  \hspace*{0.5cm}
  \includegraphics[height=0.25\textheight]{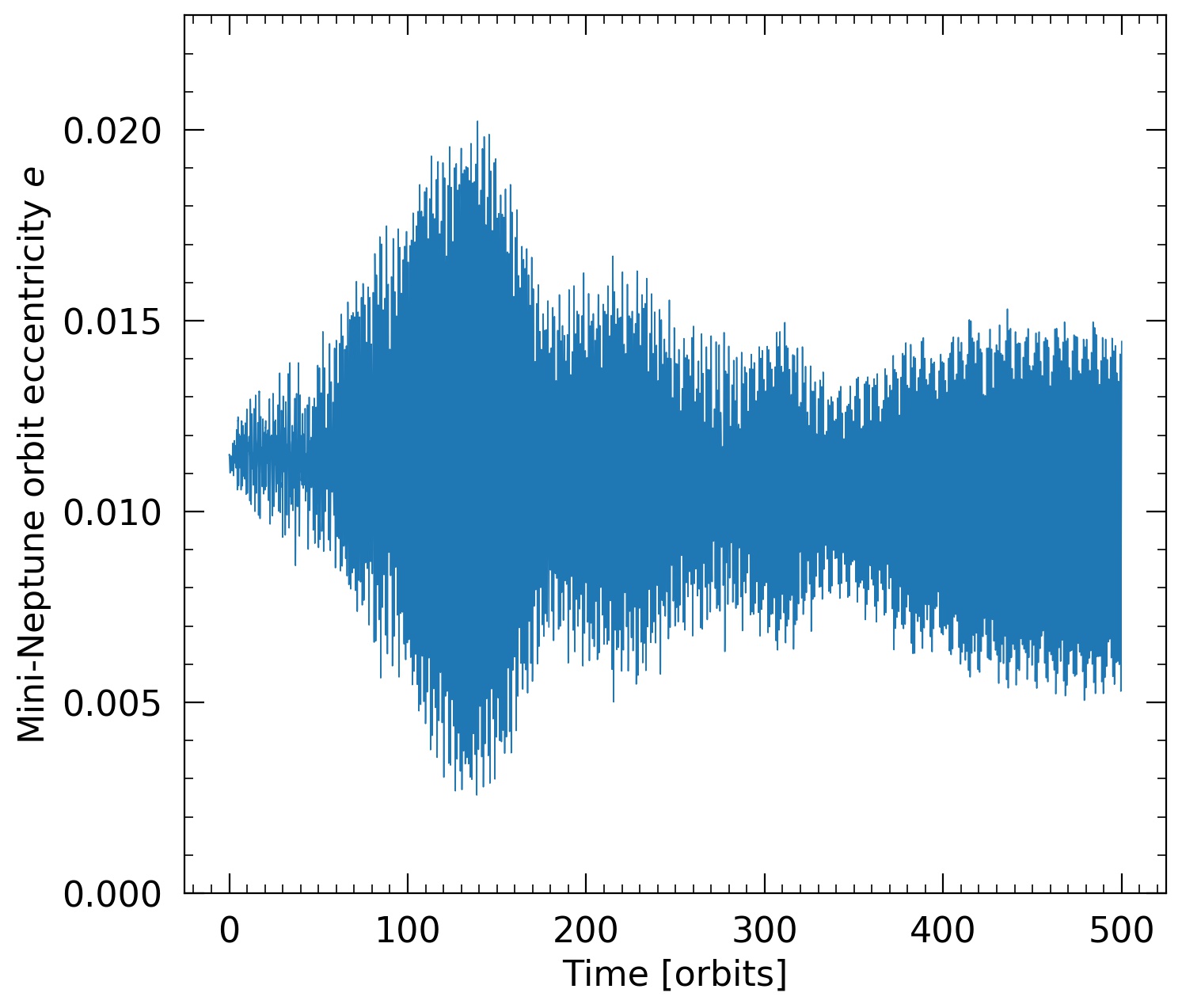}
  \caption{Orbital radius (a) and eccentricity (b) of the migrating
    Mini-Neptune in the presence of an inner giant planet. The
    amplitude of the oscillations in eccentricity and orbital radius
    reflects the impact of the giant planet's wakes on the mini
    Neptune's orbit.  The planet's initial eccentricity arises because
    of the disk's gravity being felt by the planet in addition to that
    of the star. The planet's eccentricity is calculated assuming a
    two-body problem with only the star and the planet, not the
    disc. The value of the planet’s eccentricity thus reflects the
    disc-to-star mass ratio. }
  \label{supp:orbit}
\end{figure}

Fig.~\ref{supp:giant} shows that the presence of a giant planet at
38~{\sc au} carves a wide gap in the gas at the location of the D2
dust gap. At the same time, the mini-Neptune shapes the outer region
into the triple ring system, consistent with the single planet
simulation.  The presence of the giant planet makes the morphology of
the outer rings appear eccentric.  The asymmetry in the outer rings
mutual separation is also reproduced. The migration of the
mini-Neptune is slightly affected by the gas giant's wakes (see
Fig.~\ref{supp:orbit}). Vortices develop at the outer edge of the
giant's gap but these start decaying shortly after a couple of hundred
orbits of the mini-Neptune (see right panel in
Fig.~\ref{supp:giant}). The lack of observational evidence for these
vortices may suggest that the mini-Neptune formed long after the giant
planet carved the gap D2, at a time where the vortices had already
decayed.

\begin{figure}
  {\small {a.} \hspace*{0.33\textwidth} {b.} \hspace*{0.33\textwidth}  {c.}}\\
  \includegraphics[width=0.3\textwidth]{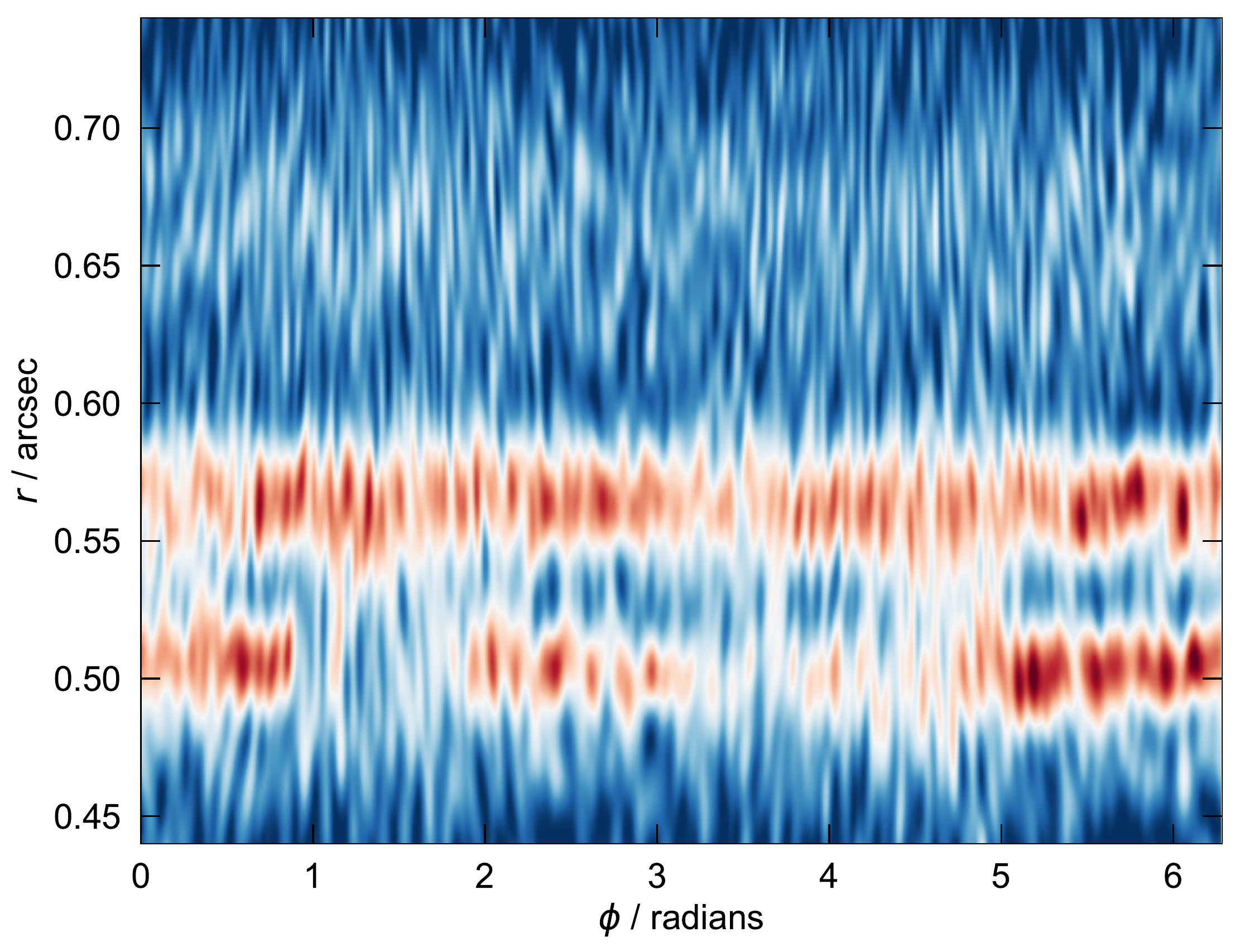}\hfill
  \includegraphics[width=0.3\textwidth]{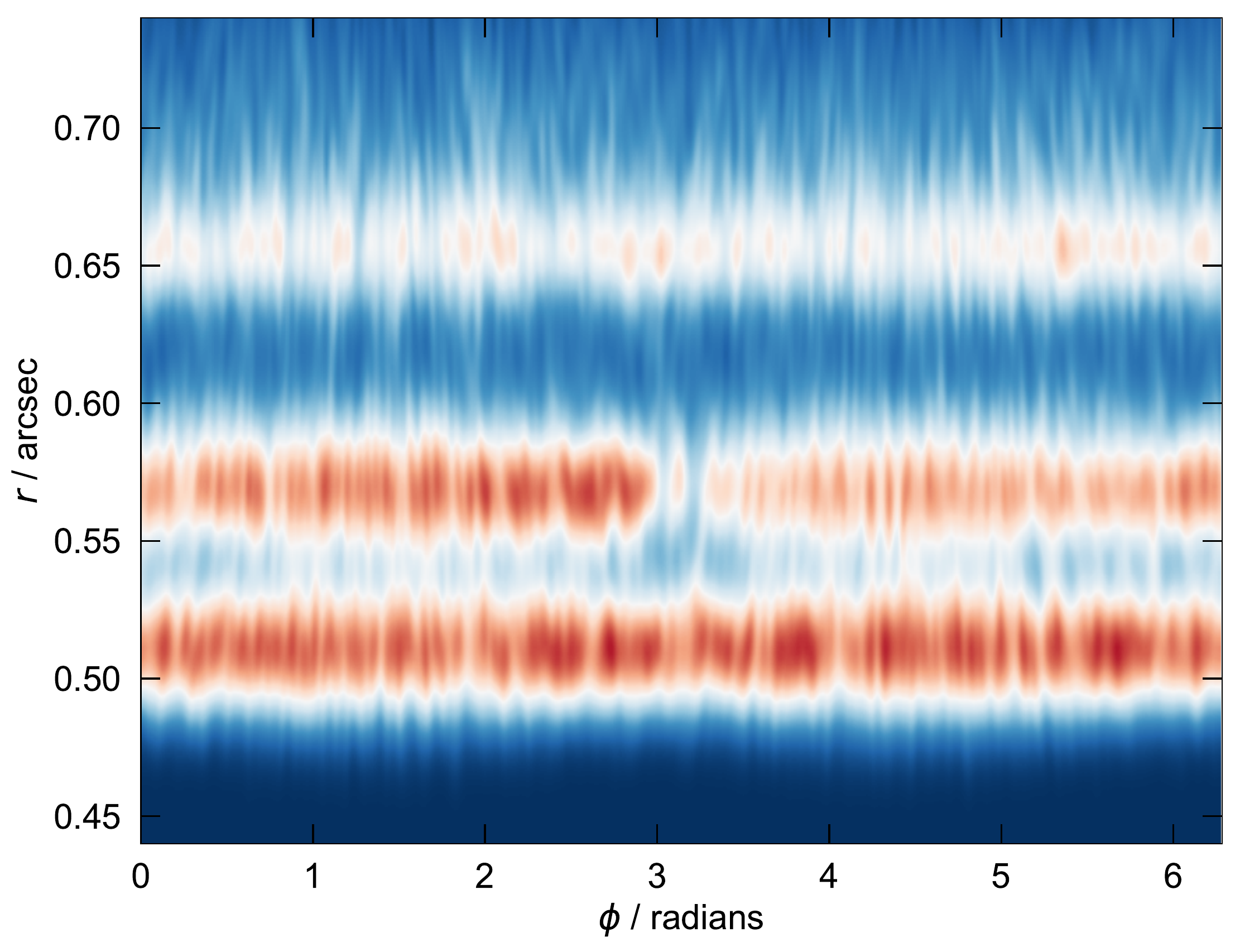}\hfill
  \includegraphics[width=0.3\textwidth]{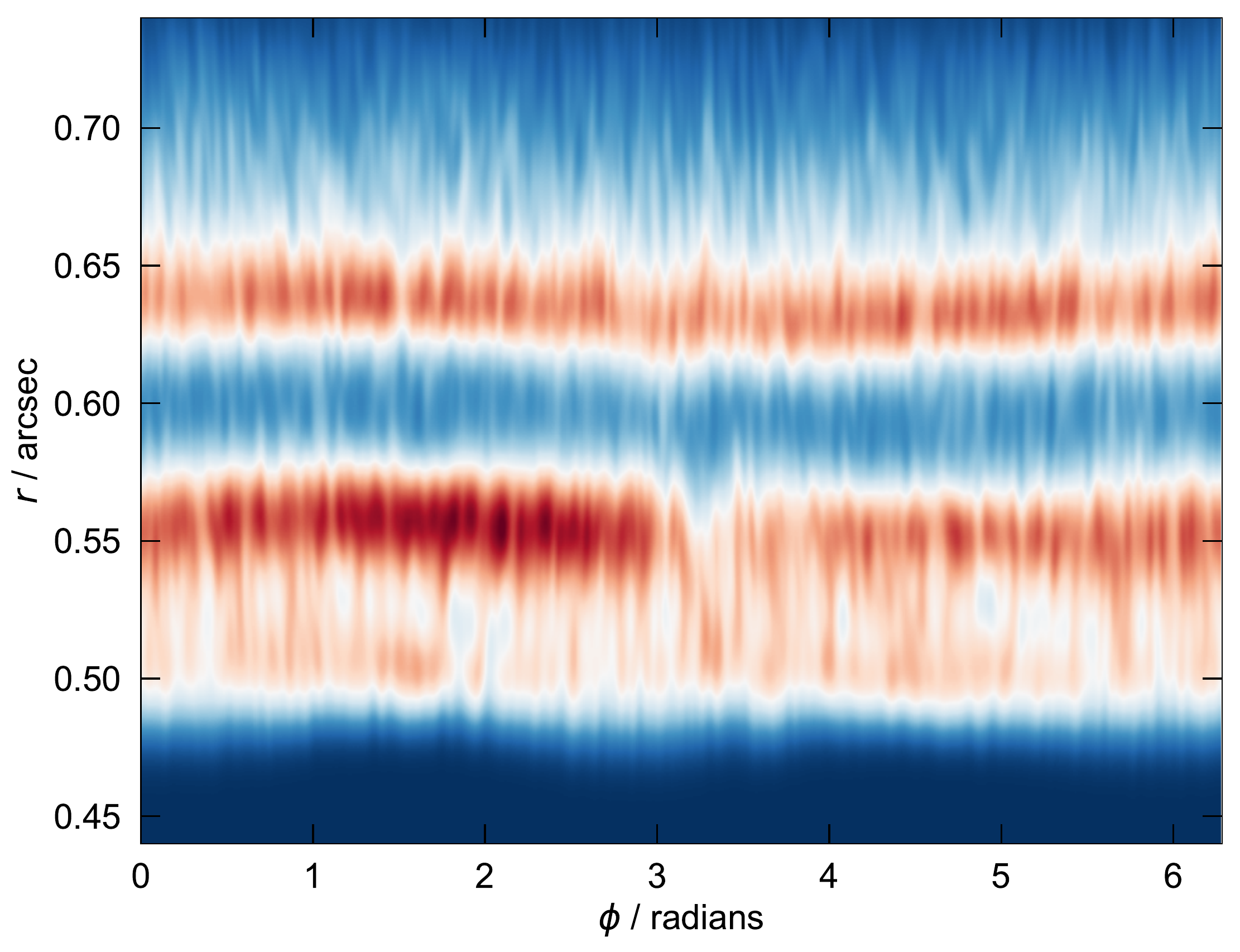}
  \caption{Comparison between polar deprojection of the ALMA
    observation (a) and synthetic predictions from hydrodynamic models
    after 150 orbits of the mini-Neptune. Panel b shows the result of
    the hydrodynamic simulations for a single mini-Neptune (same as
    Fig.~\ref{fig:model}), while panel c shows the mini-Neptune's
    outer rings under the influence of an inner giant planet at
    38~{\sc au}. All panels have linear color stretch between 0.0 and
    0.25~mJy/beam.  }
  \label{supp:depro}

\end{figure}

%% Include this line if you are using the \added, \replaced, \deleted
%% commands to see a summary list of all changes at the end of the article.
%% \listofchanges

\end{document}